\documentclass[aps,prl,showpacs,twocolumn,groupedaddress]{revtex4}
\usepackage{graphicx}
\usepackage{psfrag}

\def\bea{\begin{eqnarray}}
\def\eea{\end{eqnarray}}
\def\ba{\begin{array}}
\def\ea{\end{array}}
\def\nn{\nonumber}

\begin{document}

\title{Space-filling bearings in three dimensions} 
\author{ R. Mahmoodi Baram}
\email[]{reza@ica1.uni-stuttgart.de}
\author{H. J. Herrmann}   
\email[]{hans@ica1.uni-stuttgart.de}
\affiliation{Institute for Computational Physics, University of
Stuttgart
Pfaffenwaldring 27, 70569 Stuttgart, Germany}
\author{N. Rivier}
\email[]{nick@fresnel.u-strasbg.fr}
\affiliation{LDFC, Universit\'e Louis Pasteur, 3, rue de
l'Universit\'e F 67084 Strasbourg
cedex}

\date{\today}
\begin{abstract}
We present the first space-filling bearing in three
dimensions. It is shown that a packing which contains only
loops with even number of spheres can be constructed in a self-similar way
and that it can act as 
 a three dimensional bearing in which
spheres can rotate without slip and with negligible torsion friction.
\end{abstract}
\maketitle

Space-filling bearings have been introduced in several contexts, such as in
explaining the so-called seismic gaps \cite{McCann,Roux} of geological
faults in which two tectonic 
plates slide against each other with a friction much less than the
expected one, without production of earthquakes or of any significant heat. 
Space-filling bearings have also been used as toy models for turbulence and
can also be used in mechanical devices \cite{Grisard}. Two dimensional
space-filling bearings have been shown to exist and a discrete infinity
of
realizations has been constructed \cite{hans-prl,hans-gen}. The remaining 
question still open is: Do they also exist in three dimensions? 
This question is of fundamental importance to the physical applications.

In this letter, we will report the discovery of a self-similar
space-filling bearing in which an arbitrary chosen sphere  
can rotate around 
any axis and all the other spheres rotate accordingly 
without any sliding and with negligible torsion friction.

In two dimensions, different classes of space-filling bearings of disks have been constructed
in Refs. \cite{hans-prl,hans-gen} by requiring the loops to
have an {\em even} number of disks, since in two dimensions this is obviously 
a necessary and sufficient 
condition for disks to be able to rotate without 
any slip. Successive disks must rotate, in alternation,  
clockwise and counter-clockwise in order to avoid frustration.

\begin{figure}
\includegraphics[width=0.47\textwidth]{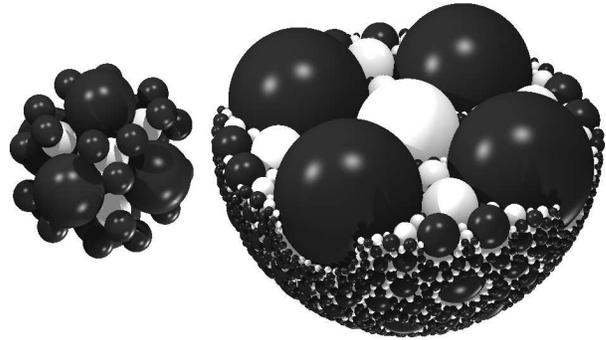}
\caption{\label{bichromatic} Bichromatic packing of spheres with octahedral
symmetry. No two spheres
of same color touch each other. The image on the left shows the initial
configuration and the first generation of inserted spheres.}
\end{figure}

The situation in three dimensions is different from two dimensions in two
ways; The axes of rotation need not be parallel, and the centers of
spheres in a loop may not lie all in the same plane. As a result,
even in an {\em isolated} odd loop spheres could rotate
without friction. But, as we will see, in the packings with an infinite number
of interconnecting loops, we could construct unfrustrated configurations of rotating spheres 
when all loops have an even number of spheres. Such a packing is   
{\em bichromatic}, i.e., one can color the spheres using only two different
colors in such a way that no spheres of same color touch each other, as shown in
Fig.\ref{bichromatic}.

No three dimensional space-filling bearing has been known up to now.
The classical Apollonian packing is space-filling and self-similar but not
a bearing since at least {\em five} colors are needed to assure different colors
at each contact. This packing can be constructed in different
ways \cite{Peikert1994,Boyd}.
By generalizing the inversion technique used in Ref.\cite{Peikert1994} to other
{\em Platonic Solids} than the tetrahedron (the base of 3D Apollonian packing) we were able
to construct five new packings including a bichromatic one. Details on the
construction algorithm and on the complete set of new configurations will be
published elsewhere \cite{paper1}. We give here only a 
qualitative description of this technique for the bichromatic packing.

Let us consider filling a sphere of unit
radius. The filling procedure is initialized by placing seven initial spheres on the
vertices and the center of a regular {\em octahedron} inside the unit
sphere, so that the spheres on the
vertices do not touch each other but touch the unit sphere and 
the one in the center. Further spheres are
inserted by inversion \cite{Mandelbrot} such that this topology can be preserved on all scales, imposing
that, all
spheres are on vertices or centers of (deformed) octahedra. 

The inversions are made iteratively with respect to nine {\em inversion
spheres}\footnote{The inversion spheres are used for the construction of the
packing and are not a part of it.}:
 One inversion sphere is concentric with the unit sphere, and is
perpendicular to the six initial spheres on the vertices of the octahedron
\footnote{Two intersecting spheres are called perpendicular if
their tangent planes on the circle of intersection are perpendicular.}. 
Inversion around it, therefore, leaves the vertex spheres invariant, and  maps the unit
sphere onto the central sphere. The other eight inversion spheres are
associated to each face of the octahedron, that is,
they are perpendicular to the unit sphere and to the three spheres on
the vertices of that face. The inversion around each of these inversion spheres maps
the unit sphere and spheres of the corresponding face into themselves (that
is, it gives no new sphere), 
and the other four initial spheres into four new
spheres within the space between the face and the unit sphere. 

\begin{figure}
\includegraphics[width=0.47\textwidth]{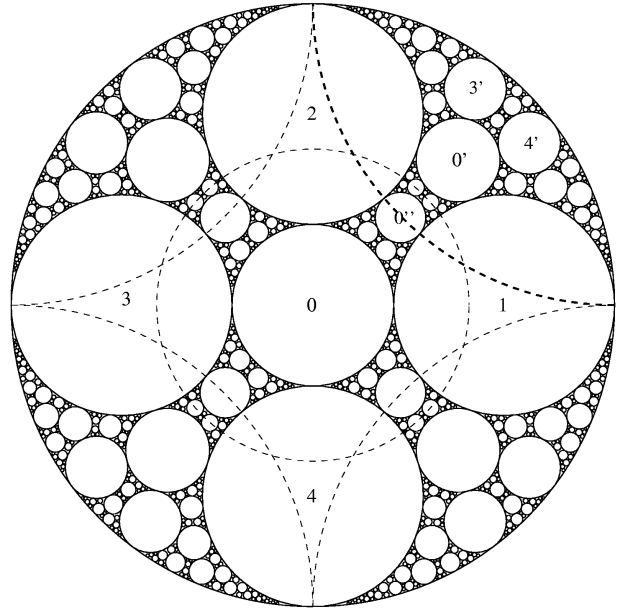}
\caption{\label{cut} Plane cut through the center of the unit sphere and
four initial spheres. Dashed circles are cuts of the inversion spheres. Sphere
$S$ is mapped, by the inversion sphere shown by a thick dashed line, onto sphere
$S'$. $0'$ is mapped by the central inversion sphere onto $0''$. }
\end{figure}

Figure \ref{cut} shows the plane cut through the centers of the unit sphere and four
initial spheres. Dashed circles are cuts of the inversion spheres. Sphere
$S:0,1,\cdots$ is mapped, by the inversion sphere shown by a thick dashed line, onto sphere
$S'$. The inversion around this sphere gives no new images of spheres $1$
and $2$.  In the first iteration we make all possible inversions
which give new and smaller spheres. In the next
iterations, the newly-generated spheres are mapped to smaller spheres. For example,
sphere $0'$ is mapped (by the central inversion sphere) onto $0''$. In this way, the 
remaining empty space is filled in the limit of infinite iterations 
while the bichromatic topology of the contacts is preserved. 

Using this algorithm, the configuration of initial spheres which gives a
bichromatic packing is unique. Strictly speaking, it is shown that the only value for
radii of the initial spheres which
leads to the bichromatic packing without (partial) overlapping of generated
spheres is  $(\sqrt{3}-1)/2$ and only using an octahedral base. The fractal
dimension of the packing has been computed using the same  method as
in Ref.\cite{hans-prl} and it is $2.59$ considerably higher than that
of Apollonian packing in 3D, that is 2.47\cite{Peikert1994}.

The image on the left of the Fig.\ref{bichromatic} shows the initial
configuration and the first generation of inserted spheres, and the one on
the right shows the resulting packing containing all the spheres with radii 
greater than $2^{-7}$. The sphere at the center and the external hull are
white  
and those on the vertices of the octahedron are black. Since the spheres on the
vertices touch only the external hull and the central sphere, and since this
topology is preserved by construction, no two 
spheres of the same color touch each other.

This implies that every loop of spheres in this packing contains an even number
of spheres. We now show that this is a sufficient condition for the
spheres in contact to rotate without slip, or even torsion friction.

\begin{figure}
\begin{center}
\psfrag{OM1}[c][c][1]{$\vec\omega_1$}
\psfrag{OM2}[c][c][1]{$\vec\omega_2$}
\psfrag{OM3}[c][c][1]{$\vec\omega_3$}
\psfrag{OM4}[c][c][1]{$\vec\omega_4$}
\psfrag{R12}[c][c][.9]{$\vec R_{12}$}
\includegraphics[width=0.45\textwidth]{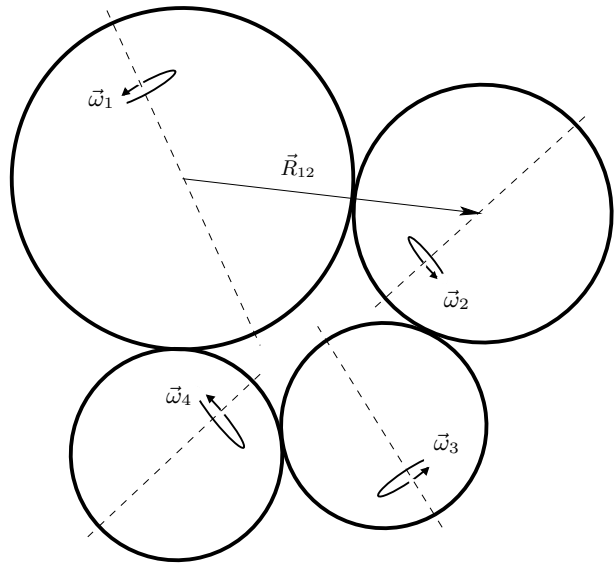}
\caption{\label{rolling} Schematic configuration of a loop of four spheres. 
The spheres rotate without slip. The centers of spheres need not be
in the same plane, although $\vec\omega_1$, $\vec\omega_2$ and
$\vec R_{12}$ are coplanar.}
\end{center}
\end{figure}

Consider a loop of $n$ spheres as seen schematically for $n=4$ in
Fig.\ref{rolling}. 
The no-slip condition implies that each pair of touching spheres have the same tangent
velocities $\vec v$ at their contact point. The condition for the contact
between the first and the second sphere can be written as:
\bea
&&\vec v_1=\vec v_2 \nn\\
\Rightarrow&&R_1\hat r_{12}\times\vec\omega_1=-R_2\hat
r_{12}\times\vec \omega_2\nn\\
\Rightarrow&&(R_1\vec \omega_1+R_2\vec \omega_2)\times \hat r_{12}=0,
\label{slipless-con}
\eea
where $R_1$, $R_2$, $\vec\omega_1$ and $\vec\omega_2$ are the radii and
the vectorial angular velocities of the first and second sphere, respectively.
$\hat r_{12}$ is the unit vector in the direction connecting the centers of the first 
and the second sphere. From Eq.(\ref{slipless-con}) the vector 
$(R_1\vec \omega_1+R_2\vec \omega_2)$ 
should be parallel to $\hat r_{12}$:  
\bea
R_2\vec \omega_2=-R_1\vec \omega_1-\alpha_{12} \hat r_{12},\label{contact12}
\eea
where $\alpha_{12}$ is an arbitrary parameter. Equation (\ref{contact12}) is a
 connection between the rotation vectors $\vec\omega_1$ and $\vec\omega_2$ of
the two spheres in contact. Similarly for the third sphere in contact with the 
second, we have 
\bea
R_3\vec \omega_3=-R_2\vec \omega_2-\alpha_{23} \hat r_{23}.\label{contact23}
\eea
Putting Eq.(\ref{contact12}) into Eq.(\ref{contact23}) we find the relation between
the angular velocities of the first and third sphere:
\bea
R_3\vec \omega_3=R_1\vec \omega_1+
\alpha_{12} \hat r_{12}-\alpha_{23} \hat r_{23}\label{contact13}. 
\eea
In general, we can relate the angular velocities of the first
and $j$th spheres  of an arbitrary chain of spheres in no-slip contacts by:
\bea
R_j\vec \omega_j=(-1)^{j-1} R_1\vec\omega_1+\sum^{j-1}_{i=1} (-1)^{j-i}
\alpha_{i,i+1}
\hat r_{i,i+1}.\label{contact1j}
\eea
As long as the chain is open, the spheres can rotate without
slip with the angular velocities given by Eq.(\ref{contact1j}) and
no restrictions on $\alpha_{i,i+1}$. 
But, for a loop of $n$ spheres in contact, spheres $j$ and $j+n$ are
identical, so that

\bea
R_1\vec \omega_1=(-1)^{n} R_1\vec\omega_1+\sum^{n}_{i=1} (-1)^{n-i+1}
\alpha_{i,i+1}
\hat r_{i,i+1}.\label{contact1n+1}
\eea 
A similar equation holds for every sphere $j = 1,\cdots,n$ in the loop.

Although for a single loop there are many solutions of
Eq.(\ref{contact1n+1}), not all will serve our purpose. 
In a packing,  each sphere belongs to a very large number of loops and
all loops should be consistent and avoid frustration.
In other words, the angular velocity obtained for a sphere as a member of
one loop should be the same as being a member of any other loop. 

If the loop contains an even number $n$ of spheres, Eq.(\ref{contact1n+1})
becomes a relation between the hitherto arbitrary coefficients of connection
$\alpha_{i,i+1}$,
\bea
\sum^{n}_{i=1} (-1)^{i}
\alpha_{i,i+1}
\hat r_{i,i+1}=0.\label{sum-even}
\eea
Using the fact that the loop is geometrically closed: 
\bea
\sum^{n}_{i=1} (R_i + R_{i+1})
\hat r_{i,i+1}=0,\label{geom-closed-loop} 
\eea
a solution for Eq.(\ref{sum-even}) is 
\bea
\alpha_{i,i+1}=c(-1)^i (R_i+R_{i+1}), \label{alphas}
\eea
where $c$ is an arbitrary constant. Putting this in 
Eq.(\ref{contact1j}), yields the angular velocities 
\bea
\vec \omega_j=\frac{1}{R_j}(-1)^{j}\left(-R_1\vec\omega_1+c
\vec R_{1j}\right) \label{final-con}, 
\eea
where $\vec R_{1j}$ is the vector which connects the centers of the first
and  $j$th sphere. 
As can be seen, the angular velocities only depend on the
positions of the spheres, so that the consistency between different loops can be 
automatically fulfilled providing that the parameter $c$ is the same for every
loop of the entire packing. 

In Eq.(\ref{final-con}), all the angular velocities are calculated
from $\vec\omega_1$ and $c$, which can be chosen arbitrarily. ($c=0$ corresponds
to the case when all angular velocities are parallel.)

The no-slip condition (\ref{contact12}) then reads
\bea
R_1\vec \omega_1+R_2\vec \omega_2=c\vec R_{12},\label{contact12even}
\eea
so that the vectors $\vec\omega_1$, $\vec\omega_2$ and $\vec R_{12}$ are coplanar 
(the plane of Fig.3, containing the two centers and the point of contact $A$). 
They are in general not collinear.

\begin{figure}
\begin{center}
\psfrag{OM_1}[c][c][1]{$\vec\omega_1$}
\psfrag{OM_2}[c][c][1]{$\vec\omega_2$}
\psfrag{SS_1}[c][c][.9]{$S_1$}
\psfrag{SS_2}[c][c][.9]{$S_2$}
\psfrag{C_1}[c][c][.8]{$C_1$}         
\psfrag{C_2}[c][c][.8]{$C_2$}
\psfrag{AA}[c][c][.8]{$A$}
\psfrag{R_12}[c][c][.8]{$\vec R_{12}$}
\includegraphics[width=0.45\textwidth]{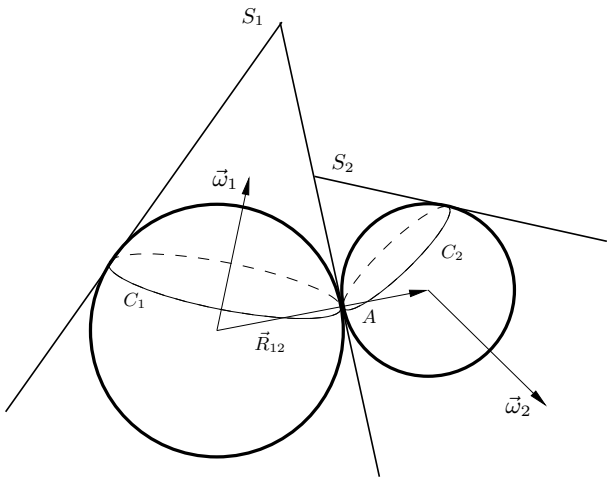}
\caption{\label{cones} Two spheres rolling on each other without slip can
be represented by two
cones rolling on each other around the common line $S_1 S_2
A$. Therefore there is no twist of the spheres against each
other.}
\end{center}
\end{figure}

We note that the condition of rotation without slip (\ref{contact12}) also
guarantees that there will be no torsion friction, as long as the three vectors are not
collinear. Indeed, the locus of the contact point $A$ on sphere $1$ is a
circle $C_1$, perpendicular to $\vec\omega_1$. On sphere $2$, it is a circle
$C_2$, perpendicular to $\vec\omega_2$. The cone tangent to sphere $i =
1,2$ on circle $C_i$ has an apex $S_i$, on the axis $\vec\omega_i$. The two
apices $S_i$ are
in the same plane as the sphere centers and on the same line as the contact point $A$ (Fig.\ref{cones}).
The two spheres rolling on each other can therefore be replaced by the two
cones rolling on each other, around the line $S_1 S_2 A$, which always
contains the contact point $A$. There
will be no twisting between cones, thus no torsion friction from one sphere rolling
on the other. Only when the two angular velocities and $\vec R_{12}$  are
collinear can there be some twist of the spheres against each other. (The
circles $C_i$ then reduce to the contact point $A$, there are no tangent cones,
and the tangent velocities $v_i = 0$). This situation is not generic. It is
of measure zero and physically irrelevant.

In the case of odd loops, Eq.(\ref{contact1n+1}) becomes 
\bea
\sum^{n}_{i=1} (-1)^{i}    
\alpha_{i,i+1}
\hat r_{i,i+1}=2R_1\vec\omega_1.\label{sum-odd}
\eea
A similar equation holds for every sphere $j = 1,\cdots,n$ in the loop. But,
since the coefficients $\alpha_{i,i+1}$ depend then on both geometry
and the rotation velocities of the spheres of the loop, consistency between
different loops cannot be fulfilled in general and, therefore, a packing
containing odd loops cannot
be a bearing. A rigorous proof for this, however, is missing. It should be
mentioned however that an unfrustrated {\em single} odd loop is possible,
but cannot occur in isolation in a packing \cite{Rivier}.

The bearing discussed here is very idealized and based on exactly-spherical
particles with infinite rigidity and, of course, does not exist
in this form on all length scales in nature. Nevertheless, chances still
remain that similar bearings, though with some imperfections, occur in reality.
A simulation of two-dimensional
shear bands shows formation of
spontaneous rotating bearings in clusters of up to 30
particles \cite{shear-band}. Despite of having more volume, the bearing
state is favored because of its low friction. As an another
evidence, the bichromatic packing presented here is self-similar which is also
observed in the samples of tectonic gouge down to several scales (see
Ref.\cite{sammis}). Interestingly, the measured fractal dimension, $2.60\pm0.1$,
agrees with that we obtained in this work.

In summary, we proved the existence of the three dimensional space-filling sphere bearing by
presenting the explicit construction.  We have shown that a 
sufficient condition is that the packing is bichromatic, and given an
explicit expression (\ref{final-con}) for the angular velocity of every sphere of the
entire packing, in terms of $\vec\omega_1$ and $c$ only. In this way, we support
the previous modelization for lubrication between tectonic plates. This
result can also be important in mechanics and
hydrodynamics.

We acknowledge the hospitality of the Max Planck Institute for the
Physics of Complex Systems, GEOMES program, oct. 2002.


\end{document}